\documentstyle [twocolumn,epsf]{mn}
\newcommand{\mincir}{\raise
-3.truept\hbox{\rlap{\hbox{$\sim$}}\raise4.truept\hbox{$<$}\ }}
\newcommand{\magcir}{\raise
-3.truept\hbox{\rlap{\hbox{$\sim$}}\raise4.truept\hbox{$>$}\ }}
\newcommand{\minmag}{\raise
-3.truept\hbox{\rlap{\hbox{$<$}}\raise5.truept\hbox{$<$}\ }}
\newcommand{\be}{\begin{equation}}
\newcommand{\ee}{\end{equation}}
\newcommand{\ba}{\begin{eqnarray}}
\newcommand{\ea}{\end{eqnarray}}
\newcommand{\brr}{\begin{array}}
\newcommand{\err}{\end{array}}
\newcommand{\bc}{\begin{center}}
\newcommand{\ec}{\end{center}}

\renewcommand{\baselinestretch}{1.0}

\title[Constraints on Cosmological and Biasing models using AGN clustering]
{Constraints on Cosmological and Biasing models using AGN clustering}
\author[Basilakos, S.]{Spyros Basilakos$^{1}$. \\
\vspace{0.1cm}
$^1$ Astrophysics Group, Imperial College London, Blackett Laboratory, 
Prince Consort Road, London SW7 2BW, UK\\
}

\begin{document}

\maketitle

\begin{abstract}
We attempt to put constraints on different cosmological and biasing
models by combining the recent clustering results of X-ray sources in
the local ($z\le 0.1$) and distant universe ($z\sim 1$).
To this end we compare the measured angular 
correlation function for bright (Akylas et al. 2000) and faint 
(Vikhlinin \& Forman 1995) {\em ROSAT} X-ray sources respectively
with those expected in three spatially flat cosmological models. Taking
into account the different functional forms of the 
bias evolution, we find that there are two cosmological models which performs well the data. 
In particular, low-$\Omega_{\circ}$ 
cosmological models ($\Omega_{\Lambda}=1-\Omega_{\circ}=0.7$) which contain either (i) high 
$\sigma_{8}^{\rm mass}=1.13$ value with galaxy merging
bias, $b(z) \propto (1+z)^{1.8}$ or (ii) low 
$\sigma_{8}^{\rm mass}=0.9$ with non-bias, $b(z) \equiv 1$ best reproduce the 
AGN clustering results. While $\tau$CDM models with different bias behaviour 
are ruled out at a high significance level.

\end{abstract}

\vspace{0.2cm}

\section{Introduction}

The study of the distribution of matter on
large scales, based on different extragalactic objects,
provides important constraints on models of cosmic structure formation.
In particular Active Galactic Nuclei (AGN) can be detected up to very high redshifts
and therefore provide information on how the X-ray selected sources trace the 
underlying mass distribution as well as the evolution of large scale structure  
(cf. Hartwick \& Schade 1989). 

However, a serious problem here is how the luminous matter traces the 
underlying mass distribution. Many authors have claimed that the large scale 
clustering pattern of different mass tracers (galaxies or clusters) 
is characterized by a bias 
picture (cf. Kaiser 1984). In this framework, biasing is assumed to be statistical in nature;
galaxies and clusters are identified as
high peaks of an underlying, initially Gaussian, random density field. 
Biasing of galaxies with respect to the dark matter distribution was also 
found to be an essential ingredient of CDM models of galaxy formation in 
order to reproduce the observed galaxy 
distribution (cf. Benson et al. 2000). 
Furthermore, different studies have shown that the bias factor, $b(z)$,
is a monotonically increasing function of redshift (cf. Fry 1996; Mo \& White 1996; 
Matarrese et al. 1997; Moscardini 
et al. 1998; Tegmark \& Peebles 1998; Basilakos \& Plionis 2001). 
For example, Steidel et al. (1998) confirmed that the 
Lyman-break galaxies are 
very strongly biased tracers of mass and they found that 
$b(z=3.4) \magcir 6, 4, 2$, for SCDM, $\Lambda$CDM $(\Omega_{\circ}=0.3)$ and
OCDM $(\Omega_{\circ}=0.2)$, respectively. 

Studies based on the traditional indicators of clustering, like the 
two point correlation function, have been utilized in order to 
describe the AGN clustering properties. Our knowledge
regarding the AGN clustering comes mostly
from optical surveys for QSO's (cf. Shanks \& Boyle 1994; Croom \& Shanks 1996; 
La Franca et al. 1998 and reference therein).  
It has been established, from Croom \& Shanks (1996), that QSO's have 
a clustering length of $r_{\circ}=5.4\pm 1.1h^{-1}$Mpc 
(with mean redshift of 1.27), while La Franca et al. (1998), analysing 
a sample of $\sim 700$ quasars
in the redshift range $0.3\le z \le 3.2$, found $r_{\circ}=6.2\pm 1.6h^{-1}$Mpc. 
It is very important to note that comparison of these clustering 
results in different redshifts 
rather favours a comoving model for the evolution of clustering.

Similarly, Vikhlinin \& Forman (1995) studied the 
angular clustering properties using a set of deep {\em ROSAT} observations.
Carrera et al. (1998) combined two soft X-ray surveys (235 AGN), 
the {\em ROSAT} Deep Survey (Georgantopoulos et al. 1996) 
and the RIXOS survey (Mason et al. 2000) and found a spatial correlation 
length $1.5h^{-1}$Mpc $\le r_{\circ} \le 5.5h^{-1}$Mpc, 
depending on the adopted model of clustering evolution. Boyle \& Mo (1993) analysing 
the EMSS survey which contains 183 low redshift AGNs, found a feeble 
clustering signal on scales $<10h^{-1}$Mpc. 
Recently, Akylas et al (2000) using  
2096 sources detected in the {\em ROSAT} All Sky Survey Bright Source Catalogue (RASSBSC), 
derived the AGN angular correlation function in the nearby Universe 
and utilizing Limber's equation obtained 
$r_{\circ}=6.7\pm 1.0h^{-1}$Mpc, assuming comoving clustering evolution.  

In this paper we present the standard theoretical approach  
to estimate the angular correlation 
function $w(\theta)$, using different models for the bias evolution in 
different spatially flat cosmological models.
Comparing the latter with observational results, 
we attempt to put constraints on the different cosmological and bias
models. 
The plan of this paper is the following:
In section 2, (a) we present the calculation of the theoretical predictions for $w(\theta)$, 
(b) we describe the models for bias evolution and (c) we discuss the AGN selection function. 
In section 3 we present the CDM spatially flat cosmologies, while in section 4 we present the 
predictions for $w(\theta)$. The observational results are compared with the predictions in section
5 and finally in section 6 we draw our conclusions.

\section{The Integral Equation}
For the purpose of this study we will utilize the relation between the angular 
$w(\theta)$ and spatial $\xi(r,z)$ two point 
correlation functions (cf. Maglioccheti et al. 1999 and references theirin).
As it is well known, this connection can be done using the Limber 
equation (cf. Peebles 1980). For example, in the
case of a spatially flat Universe ($\Omega_{\circ}+\Omega_{\Lambda}=1$), the Limber 
equation can be written as
\be 
w(\theta)=2\frac{\int_{0}^{\infty} \int_{0}^{\infty} x^{4} \phi^{2}(x) \xi(r,z) {\rm d}x {\rm d}u}
{[\int_{0}^{\infty} x^{2} \phi(x){\rm d}x]^{2}} \;\; , 
\ee
where $\phi(x)$ is the selection function (the probability 
that a source at a distance $x$ is detected in the survey) and 
$x$ is the comoving coordinate related to the redshift through 
\be
x=\frac{c}{H_{\circ}} \int_{0}^{z} \frac{{\rm d}y}{E(y)} \;\; ,
\ee
with 
\be 
E(z)=[\Omega_{\circ}(1+z)^{3}+\Omega_{\Lambda}]^{1/2} 
\ee
(see Peebles 1993). The mean surface density, ${\cal N}$,
on a survey of solid angle $\Omega_{s}$ is:

\be
{\cal N}=\int_{0}^{\infty} x^{2} \phi(x) {\rm d}x=\frac{1}{\Omega_{s}} 
\int_{0}^{\infty} N(z) {\rm d}z \;\; ,
\ee
where $N(z)$ is the number of objects in the given survey within the shell $(z,z+{\rm d}z)$. 
Therefore, combining the above system of 
equations, the expression for $w(\theta)$ satisfies the form 

\begin{equation}\label{eq:angu}
w(\theta)=2\frac{H_{\circ}}{c} \frac{\int_{0}^{\infty} N^{2}(z)E(z){\rm d}z 
\int_{0}^{\infty} \xi(r,z) {\rm d}u}{[\int_{0}^{\infty} N(z) {\rm d}z]^{2}} \;\;.
\end{equation} 
The physical separation between two sources, 
separated by an angle $\theta$ considering 
a small angle approximation is given by:

\be
r \simeq \frac{1}{(1+z)} \left( u^{2}+x^{2}\theta^{2} \right)^{2} \;\; .
\ee
Extending this picture, 
we quantify the evolution of clustering with epoch 
presenting the spatial correlation function of the X-ray sources as  
\be
\label{eq:spat}
\xi(r,z)=\xi_{\rm mass}(r)R(z) \;\;, 
\ee
with 
\be
\label{eq:spat1}
R(z)=D^{2}(z)b^{2}(z)
\ee
where $D(z)$ is the linear growth rate of clustering (cf. Peebles 1993)
\footnote{$D(z)=(1+z)^{-1}$ for an Einstein-de Sitter Universe.} being given by   
\be\label{eq:24}
D(z)=\frac{5\Omega_{\circ} E(z)}{2}\int^{\infty}_{z} \frac{(1+y)}{E^{3}(y)} 
{\rm d}y 
\ee
and $b(z)$ is the 
evolution of bias. 
On large scales, we utilize the standard procedure based on the linear theory 
predictions from the power spectrum $P(k)$. On small scales we will 
follow the notation of Maglioccheti et al. (1999), who used
the APM correlation function (Maddox et al. 1990) in 
order to estimate the angular correlation function of radio sources.
In partcular, for $\Gamma=0.2$ the above extrapolation is used for 
$r \le r_{\circ}^{\rm opt}=5.4 h^{-1}$Mpc while for other values of $\Gamma$ we have 
re-scaled $\xi(r)$ using $\xi(r) \propto \xi(r \Gamma_{1}/\Gamma_{2})$.

\subsection{Bias Evolution}
The concept of biasing between different classes of extragalactic objects 
and the background matter distribution was put forward by Kaiser (1984)
and Bardeen et al. (1986) 
in order to explain the higher amplitude of the 2-point correlation function 
of clusters of galaxies with respect to that of galaxies themselves.

The deterministic and linear nature of 
biasing has been challenged (cf. Bagla 1998; Dekel \& Lahav 1999) and indeed 
on small scales ($r< 10h^{-1}$Mpc) there are significant deviations from $b(r)=const$. 
Despite the above, the linear biasing assumption is still a useful first order
approximation which, due to its simplicity, it is used in most studies
of large scale clustering (cf. Magliocchetti et al. 1999). 
In this paper however, we will not indulge any variation of bias with scale
but rather, working within the paradigm of linear and 
scale-independent bias. Therefore, we shortly describe 
some of the bias evolution models in order to introduce them to our analysis
(see eq. \ref{eq:spat} and eq. \ref{eq:spat1}).

\begin{itemize}
\item No evolution of bias (B0): This model considers constant bias at all 
epochs: 
\be
\label{eq:pres}
b(z)=b_{\circ} \simeq b_{\circ}^{\rm opt} \left( \frac{r_{\circ}}{r_{\circ}^{\rm opt}} \right)^{\gamma/2}
\ee
where $r_{\circ}^{\rm opt}=5.4h^{-1}$Mpc is the correlation length in comoving coordinates 
estimated by the APM correlation function (Maddox et al. 1990) and 
$r_{\circ}$ is the corresponding length for 
the X-ray sources. Finally, $b_{\circ}^{\rm opt}=1/\sigma_{8}^{\rm mass}$ is the present bias of optical 
galaxies relative to the distribution of mass and $\sigma_{8}^{\rm mass}$ is the mass 
rms fluctuations in sphere of radius 8$h^{-1}$Mpc. Using the above ideas, 
if ones assumes that $r_{\circ} > r_{\circ}^{\rm opt}$, then it is quite 
obvious that the X-ray sources are indeed more biased with respect to optical
galaxies by the factor $(r_{\circ}/r_{\circ}^{\rm opt})^{\gamma/2}$. In case 
that $b(z)\equiv  1$, we have the so called 
non-bias model\footnote{From now on we consider $b(z)\equiv 1$ as a (B0) bias model.}.

\item  Test Particle or Galaxy Conserving Bias (B1): 
This model, proposed by Nusser \& Davis (1994), Fry (1996), Tegmark \&
Peebles (1998), predicts the evolution of bias,
independent of the mass and the origin of halos, assuming only that the
test particles fluctuation field is related proportionally
to that of the underlying mass. 
Thus, the bias factor as a function of redshift can be written:
\be
b(z)=1+\frac{(b_{\circ}-1)}{D(z)} \;\; \;\; , 
\ee
where $b_{\circ}$ is the bias factor at the present time. 
Bagla (1998) has found, that in the range $0\le z \le 1$,
the above formula describes well the evolution of bias.

\item Merging Bias Model (B2): 
Mo \& White (1996) have developed a model for 
the evolution of the correlation bias, 
using the Press-Schechter formalism.
Utilizing a similar formalism, Matarrese et al. (1997) extended the Mo \& White
results to include the effects of different mass scales (see also 
Moscardini et al. 1997; Bagla 1998; Catelan et al. 1998; Magliocchetti et al. 2000). In this case we have that

\be
b(z)=0.41+\frac{(b_{\circ} - 0.41)}{D(z)^{\beta}} \;\; ,
\ee
with $\beta \simeq 1.8$.
\end{itemize}

\subsection{Selection Function}
In flux-limited samples, it is well known that  
there is a degradation of
sampling as a function of distance from the observer (codified by the
so called {\em selection function}). The latter also depends on the evolution
of the source luminosity function. Thus for the X-ray sources the selection function 
can be written as 

\be 
\phi(x)=\int_{L_{\rm min}}^{\infty} \Phi(L_{x},z) {\rm d}L \;\;.
\ee
In this work we used a luminosity function of the form assumed by Boyle et al. (1993), 
which takes into account the cosmological evolution of the QSO's in the form of pure luminosity 
evolution.

\section{Cold Dark Matter (CDM) Cosmologies}

In this section, we present the cosmological models that we use in this work.
For the power spectrum of our CDM models, we consider $P(k) \approx k^{n}T^{2}(k)$ with
scale-invariant ($n=1$) primeval inflationary fluctuations. We utilize the transfer function 
parameterization as in Bardeen et al. (1986), with the corrections given 
approximately by Sugiyama's (1995) formula:

\vspace{1.2cm}

$$T(k)=\frac{{\rm ln}(1+2.34q)}{2.34q}[1+3.89q+(16.1q)^{2}+$$
$$(5.46q)^{3}+(6.71q)^{4}]^{-1/4} \;\; .$$
with
\be 
q=\frac{k}{\Omega_{\circ}h^{2}{\rm exp}[-\Omega_{b}-(2h)^{1/2}\Omega_{b}/\Omega_{\circ}]}
\ee
where $k=2\pi/\lambda$ is the wavenumber in units of 
$h$ Mpc$^{-1}$ and $\Omega_{b}$ is the baryon density. 
In this analysis, we have taken into account three different 
cold dark matter models (CDM) in order to isolate the effects of different parameters 
on the X-ray sources clustering predictions. 

The $\tau$CDM\footnote{This model could correspond to the decaying 
neutrino model.} and $\Lambda$CDM (see Martini \& Weinberg 2000) are approximately
COBE normalized and the latter cosmological model
is consistent with the results from Type Ia supernovae (Riess et al. 1998;
Perlmutter et al. 1999). In the same framework, the $\tau$CDM and $\Lambda$CDM models 
have $\Gamma \sim 0.2$, in approximate agreement with the shape parameter estimated from galaxy 
surveys (cf. Maddox et al. 1990; Peacock \& Dodds 1994) and 
they have fluctuation amplitude in 8 $h^{-1}$Mpc scale, $\sigma_{8}^{\rm mass}$,  
consistent with the cluster abundance, $\sigma_{8}^{\rm mass}=0.55\Omega_{\circ}^{-0.6}$
(Eke, Cole, \& Frenk 1996). 

Furthermore, in order to investigate cosmological models with high value of 
$\sigma_{8}^{\rm mass}$, we include a new model named $\Lambda$CDM2 (cf. Cole et al. 1998). 
The $\sigma_{8}^{\rm mass}$ value for the latter cosmological model is in good 
agreement with both
cluster and the 4-years COBE data with a shape parameter $\Gamma=0.25$. Therefore, it 
is quite obvious that two of our  
models ($\tau$CDM and $\Lambda$CDM)  
have the same power spectrum and geometry but different values of $\Omega_{\circ}$ 
and $\sigma_{8}^{\rm mass}$, while the 
two spatially flat, low-density CDM models ($\Lambda$CDM and $\Lambda$CDM2)
have different $\sigma_{8}$ and $\Gamma$ respectively.  
In table 1, we present the normalizations for the above 
specific cosmological models as well as the values of the AGN bias, $b_{\circ}$, at the 
present time (described by eq. \ref{eq:pres}). Thus 
it turns out that in $\Lambda$CDM and $\tau$CDM models the distribution of X-ray 
sources is ``biased'' relative to the distribution of mass; while the $\Lambda$CDM2 model is 
almost non ``biased''.

\begin{table*}
\caption[]{Small Scale Normalizations}
\tabcolsep 6pt
\begin{tabular}{ccccccc} 

Model & $\Omega_{\circ}$  
&$\Omega_{\Lambda}$& $h$ & $\sigma_{8}^{\rm mass}$ & $b_{\circ}$&$\Omega_{b}$\\ 
$\tau$CDM&1.0&0.0&0.50& 0.55&2.18&0.050\\
$\Lambda$CDM & 0.3&0.7&0.65&0.90&1.33&0.036\\
$\Lambda$CDM2 & 0.3&0.7&0.60&1.13&1.06&0.035\\

\end{tabular}
\end{table*}

\section{Theoretical Predictions}
In this section we will first present results on the 
predicted angular correlation function $w(\theta)$ (see Figure 1) estimated 
for three different CDM spatially flat cosmological models taking into account 
the functional forms of the bias evolution (B0, B1, B2) introduced in the previous section.
From figure 1, it is obvious that the amplitude of  
$w(\theta)$ increases for models with bias evolution in all 
cosmological models. 
This is to be expected due to the fact that the amplitude of 
the angular correlation function is affected by the 
expression $R(z)$, especially at high redshifts. Indeed, 
we observe that models with increasing bias as a function of redshift, 
give a stronger clustering signal (in small angular scales), relative to models
with non-bias (B0). 
For the latter bias behaviour, it is interesting to see, that the 
predictions for $w(\theta)$ for the case of 
the $\tau$CDM and $\Lambda$CDM2 are almost the same.

\begin{figure}
\mbox{\epsfxsize=9cm \epsffile{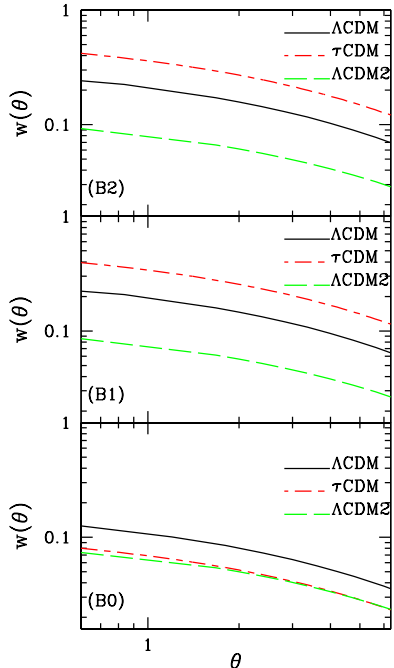}}
\caption{Theoretical predictions of the angular correlation function $w(\theta)$.} 
\end{figure}
Therefore, in order to understand better the effects of AGN clustering,
we present in figure 2 the quantity $R(z)=D^{2}(z)b^{2}(z)$ as a 
function of redshift for the three cosmological models, utilizing 
at the same time different bias evolution. It is quite
obvious that the behaviour of the function $R(z)$ characterizes the  
clustering evolution with epoch. Figure 2, for example, 
clearly shows that the bias at high redshifts has different values in different
cosmological models. In particular for the high $\sigma_{8}^{\rm mass}$ low-$\Omega_{\circ}$
flat model ($\Lambda$CDM2) 
the distribution of X-ray sources is only weakly
biased, as opposed to the strongly biased distribution in the 
$\tau$CDM cosmological model.       

Indeed the different functional forms of $b(z)$, provide clustering models 
where: 
\begin{itemize}
\item AGN clustering is a decreasing function with redshift for
(B0), 
\item AGN clustering is roughly constant for (B1). 
However, the $\Lambda$CDM2-B1 model gives lower $R(z)$
simply because the higher $\sigma_{8}^{\rm mass}$ normalization largely removes the 
clustering difference between the two other flat cosmological models
with low $\sigma_{8}^{\rm mass}$ normalizations. In other words, the present bias value 
of the above model is almost $\sim 1$, which gives clustering behaviour similar to 
the $\Lambda$CDM2-B0.
\item AGN clustering is a monotonically 
increasing function of redshift for (B2).  
\end{itemize}

\begin{figure}
\mbox{\epsfxsize=9cm \epsffile{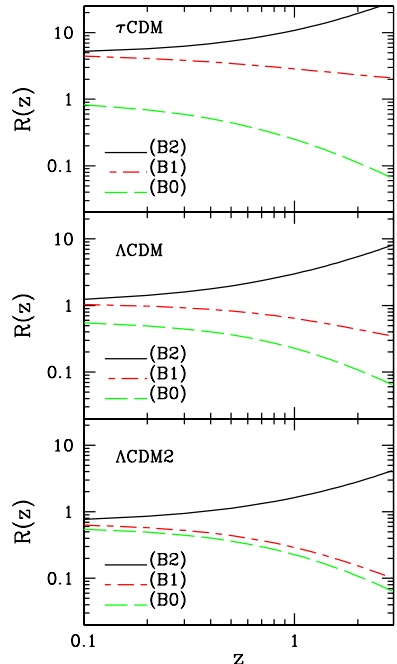}}
\caption{The function $R(z)=D^{2}(z)b^{2}(z)$ as a function of redshift, 
for different bias evolution models.} 
\end{figure}

\section{Application to the data}

In Figure 3 we compare the angular correlation function of a 
sample of 2096 sources with a total sky coverage of 4.9sr detected in the 
{\em ROSAT} All-Sky Survey Bright Source Catalogue (see Akylas et
al. 2000) with that predicted in various flat cosmological models. 
Considering a two point angular correlation function of the form
$w(\theta)=(\theta/\theta_{\circ})^{1-\gamma}$, 
the above authors
found $\theta_{\circ}=0.062^{\circ}$, $\gamma=1.8$ and spatial 
correlation length of $r_{\circ} \approx 6.5\pm 1.0h^{-1}$Mpc and 
$r_{\circ} \approx 6.7\pm 1.0h^{-1}$Mpc for stable and comoving clustering evolution 
respectively, similar to the optically selected AGN.
Due to the fact that the above estimates have been focused on the local X-ray Universe,  
in this work we use complementary observational results from Vikhlinin \& Forman (1995). They
have analysed a set of deep {\em ROSAT} observations with a total sky
coverage of 40 deg$^{2}$, 
in order to investigate the clustering 
properties of faint X-ray sources. Therefore, taking into account the correction for the 
amplification bias they claimed that the two point angular 
correlation function is well described by a power low with
$\gamma=1.8$. Thus, in figure 4 
we plot their results and the estimated angular correlation function for 
all nine models.

\begin{figure}
\mbox{\epsfxsize=9cm \epsffile{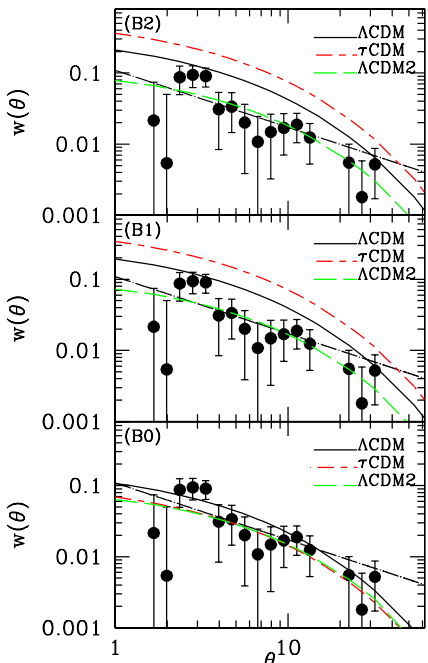}}
\caption{Comparison of the predicted angular correlation function 
for various cosmological models with that of the local AGN distribution, 
estimated by Akylas et al. (2000). 
Errorbars are determined by assuming Poisson statistics. The continuous dot-dash line
represent the best fit to $w_{x}(\theta)$ derived by the above authors.}
\end{figure}

\begin{figure}
\mbox{\epsfxsize=9cm \epsffile{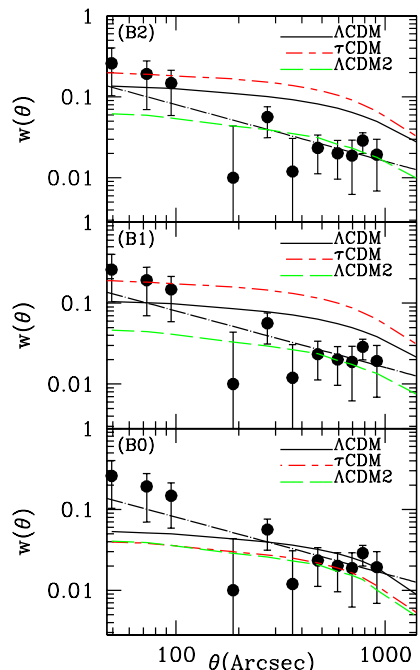}}
\caption{Comparison of the predicted angular correlation function 
for various models with that estimated by Vikhlinin \& Forman (1995). 
The continuous dot-dash line
represent the best fit to $w_{x}(\theta)$ derived by the above authors.}
\end{figure}

\begin{table}
\caption[]{$\chi^2$ probabilities (${\cal P}_{>\chi^{2}}^{B}$)
of consistency between RASSBSC data and
models.} 
\tabcolsep 8pt
\begin{tabular}{cc}
Comparison Pair &  ${\cal P}_{>\chi^{2}}^{B}$\\ 
RASSBSC - $\Lambda$CDM-B0 & 0.81 \\
RASSBSC - $\Lambda$CDM-B1 & 0.067\\
RASSBSC - $\Lambda$CDM-B2 & 0.019 \\ \\

RASSBSC - $\Lambda$CDM2-B0 & 0.19\\
RASSBSC - $\Lambda$CDM2-B1 & 0.42 \\
RASSBSC - $\Lambda$CDM2-B2 & 0.57\\ \\

RASSBSC - $\tau$CDM-B0 & 0.19 \\
RASSBSC - $\tau$CDM-B1 & 2.88$\times 10^{-8}$ \\
RASSBSC - $\tau$CDM-B2 & 3.00$\times 10^{-9}$ \\

\end{tabular}
\end{table}

\begin{table}
\caption[]{$\chi^2$ probabilities (${\cal P}_{>\chi^{2}}^{F}$)
of consistency between faint {\em ROSAT} data and
models.}
\tabcolsep 8pt
\begin{tabular}{cc}
Comparison Pair &  ${\cal P}_{>\chi^{2}}^{F}$\\ 
faint {\em ROSAT} - $\Lambda$CDM-B0 & 0.037\\
faint {\em ROSAT} - $\Lambda$CDM-B1 & 3.67 $\times 10^{-4}$\\
faint {\em ROSAT} - $\Lambda$CDM-B2 & 1.72 $\times 10^{-6}$\\ \\

faint {\em ROSAT} - $\Lambda$CDM2-B0 & 3.44 $\times 10^{-4}$ \\
faint {\em ROSAT} - $\Lambda$CDM2-B1 & 3.21 $\times 10^{-3}$ \\
faint {\em ROSAT} - $\Lambda$CDM2-B2 & 0.036 \\ \\

faint {\em ROSAT} - $\tau$CDM-B0 & 4.64 $\times 10^{-3}$ \\
faint {\em ROSAT} - $\tau$CDM-B1 & 1.07$\times 10^{-12}$ \\
faint {\em ROSAT} - $\tau$CDM-B2 & 4.73 $\times 10^{-15}$ \\

\end{tabular}
\end{table}

In order to quantify the
differences between models and data, we perform a standard 
$\chi^{2}$ test, for the bright (RASSBSC) and faint sources respectively,
and we present the ${\cal P}_{>\chi^{2}}^{N}$\footnote{Where $N=B,F$ for bright 
(RASSBSC) and faint X-ray sources respectively.}
results in tables 2 and 3.
Comparing the statistical results for 
both (a) high (faint sources) and (b) low (bright) redshift regimes,
we can point out that for the bright X-ray sources the
only models that are excluded by the data, at 
a relatively high significance level, are $\Lambda$CDM-B2, $\tau$CDM-B1 and $\tau$CDM-B2. 
Interestingly, for the faint objects the excluded models are 
$\Lambda$CDM-B1, $\Lambda$CDM-B2, $\Lambda$CDM2-B0, 
$\Lambda$CDM2-B1, $\tau$CDM-B0, $\tau$CDM-B1 and $\tau$CDM-B2. 
The above differences between the two kind of populations are to 
be expected simply because the cosmological 
evolution plays an important role on large scale structure clustering 
due to the fact that the high redshift objects are more biased tracers of 
the underlying matter distribution with respect to the low redshift
objects (cf. Steidel et al. 1998). 
Also, from the faint X-ray sources results we would like to point out that 
there is not a single $\tau$CDM model that fits the data.

If we make the reasonable assumption that there is no correlation between
the two X-ray populations, mostly due to the large distances
involved, we can consider the RASSBSC and faint {\em ROSAT} catalogues 
as being independent of each other. Under this assumption the previous
statistical tests can also be considered as independent.
Indeed performing a standard Kolmogorov-Smirnov test comparing the two different 
spatial correlation functions, described by Akylas et al. (2000) and 
Vikhlinin \& Forman (1995) we find that they are significantly different.
In this framework the joint (overall) probability can be given by the following expression:

\be 
P={\cal P}_{>\chi^{2}}^{B} {\cal P}_{>\chi^{2}}^{F}
\left[1-{\rm ln} ({\cal P}_{>\chi^{2}}^{B}{\cal P}_{>\chi^{2}}^{F})\right] \;\; .        
\ee
In table 4 we can see the corresponding joint probabilities for all our models. This 
overall statistical test proves that 
the $\Lambda$CDM-B0 and $\Lambda$CDM2-B2   
models fit well the observational data at a
relatively high significance level. 

\begin{table}
\caption[]{The joint probabilities $P$.}
\tabcolsep 9pt
\begin{tabular}{cc}
Fitted Models &  $P$\\ 
$\Lambda$CDM-B0 & 0.14\\
$\Lambda$CDM-B1 & 2.86 $\times 10^{-4}$\\
$\Lambda$CDM-B2 & 5.96 $\times 10^{-7}$ \\ \\

$\Lambda$CDM2-B0 & 7.00 $\times 10^{-4}$\\
$\Lambda$CDM2-B1 & 0.01 \\
$\Lambda$CDM2-B2 & 0.10 \\ \\

$\tau$CDM-B0 & 7.08 $\times 10^{-3}$ \\
$\tau$CDM-B1 & 1.41$\times 10^{-18}$ \\
$\tau$CDM-B2 & 7.29$\times 10^{-22}$ \\
\end{tabular}
\end{table}
Note that we have
tested the AGN clustering predictions using also the SCDM (COBE normalized 
from Bunn \& White 1997, with 
$h=0.5$, $\Gamma=0.5$ and $\sigma_{8}^{\rm mass}=1.22$) cosmological
model and we found that it is excluded by the data at 
a high significance level (${\cal P}_{>\chi^{2}}^{B,F} \simeq 10^{-6}-10^{-9}$).
Finally, a possible contamination of the AGN samples by stars, which can be
assumed at first order to be distributed randomly on the sky, will lower
the true amplitude of the AGN $w(\theta)$. We have crudely tested the effect
on our model comparion results of an AGN amplitude drop of $\sim 10$\% and found
no qualitative differences whatsoever.

We should conclude that the behaviour of the observed angular 
correlation function of the X-ray sources is sensitive to the different cosmologies but 
at the same time there is a strong dependence on the bias models that 
we have considered in our analysis. By separating between low and 
high redshift regimes, we obtain results being consistent with the 
hierarchical clustering scenario, in which the 
AGN's are strongly biased at all cosmic epochs 
(cf. Magliocchetti et al. 1999).

\section{Conclusions}
We have studied the clustering properties of 
the X-ray sources using the predicted angular correlation function 
for several cosmological models. We parametrize the predictions for $w(\theta)$ 
taking into account the behaviour of $b(z)$ for a non-bias model, 
a galaxy conserving bias model with $b(z) \propto (1+z)$ and  
for a galaxy merging bias model with $b(z) \propto (1+z)^{1.8}$. Utilising 
the measured angular correlation function, 
for faint and bright X-ray {\em ROSAT} sources,
estimated in Vikhlinin \& Forman (1995) 
and Akylas et al. (2000) respectively, we have compared them
with the corresponding ones predicted in three cosmological models,
namely the $\tau$CDM, $\Lambda$CDM and $\Lambda$CDM2 (with a high $\sigma_{8}^{\rm mass}$ value).  
We find that the models that best reproduce the observational results are: 

\begin{itemize}
\item $\Lambda$CDM2 model ($\Omega_{\Lambda}=1-\Omega_{\circ}=0.7$) with 
high $\sigma_{8}^{\rm mass}=1.13$ and bias evolution described by $b(z) \propto (1+z)^{1.8}$.
\item $\Lambda$CDM model ($\Omega_{\Lambda}=1-\Omega_{\circ}=0.7$) with 
low $\sigma_{8}^{\rm mass}=0.9$ and bias evolution by $b(z) \equiv 1$.
\end{itemize}

\section* {Acknowledgements}
I would like to thank 
the referee Dr. P. Schuecker as well as Manolis Plionis, 
M. Rowan-Robinson and Ioannis Georgantopoulos 
for their useful comments and suggestions. 
This work was supported by EC Network programme 'POE' (grant number HPRN-CT-2000-00138).

\end{document}